# *Q*-Learning for Robust Satisfaction of Signal Temporal Logic Specifications

Derya Aksaray, Austin Jones, Zhaodan Kong, Mac Schwager, and Calin Belta

*Abstract*— This paper addresses the problem of learning optimal policies for satisfying signal temporal logic (STL) specifications by agents with unknown stochastic dynamics. The system is modeled as a Markov decision process, in which the states represent partitions of a continuous space and the transition probabilities are unknown. We formulate two synthesis problems where the desired STL specification is enforced by maximizing the probability of satisfaction, and the expected robustness degree, that is, a measure quantifying the quality of satisfaction. We discuss that *Q*-learning is not directly applicable to these problems because, based on the quantitative semantics of STL, the probability of satisfaction and expected robustness degree are not in the standard objective form of *Q*-learning. To resolve this issue, we propose an approximation of STL synthesis problems that can be solved via *Q*-learning, and we derive some performance bounds for the policies obtained by the approximate approach. The performance of the proposed method is demonstrated via simulations.

## I. INTRODUCTION

This paper addresses the problem of controlling a system with unknown, stochastic dynamics to achieve a complex, time-sensitive task. An example is controlling a noisy aerial vehicle with partially known dynamics to visit a pre-specified set of regions in some desired order while avoiding hazardous areas. We consider tasks given in terms of temporal logic (TL) [4] that can be used to reason about how the state of a system evolves over time. Recently, there has been a great interest in control synthesis with TL specifications (e.g., [2], [3], [8], [22], [19], [12]). When a stochastic dynamical model is known, there exist algorithms to find control policies for maximizing the probability of achieving a given TL specification (e.g., [19], [17]) by planning over stochastic abstractions (e.g., [16], [1], [19]). However, only a handful of papers have considered the problem of enforcing TL specifications to a system with unknown dynamics. For example, reinforcement learning has been used to find a policy that maximizes the probability of satisfying a given linear temporal logic (LTL) formula in [5], [22], [12].

*This work was partially supported at Boston University by ONR grant number N00014-14-1-0554 and by the NSF grant numbers CMMI-1400167, NSF NRI-1426907.

D. Aksaray is with the Computer Science and Artificial Intelligence Laboratory, Massachusetts Institute of Technology, Cambridge, MA, USA. daksaray@mit.edu. A. Jones is with the Departments of Mechanical Engineering and Electrical Engineering, Georgia Institute of Technology, Atlanta, GA, USA. austinjones@gatech.edu. Z. Kong is with the Department of Mechanical and Aerospace Engineering, University of California Davis, Davis, CA, USA. zdkong@ucdavis.edu. M. Schwager is with the Department of Aeronautics and Astronautics, Stanford University, Stanford, CA, USA. schwager@stanford.edu. C. Belta is with the Department of Mechanical Engineering, Boston University, Boston, MA, USA. cbelta@bu.edu.

In contrast to existing works on reinforcement learning using propositional temporal logic, we consider signal temporal logic (STL), a rich predicate logic that can be used to describe tasks involving bounds on physical parameters and time intervals [10]. An example STL specification is "Within $t_1$ seconds, a region in which $y$ is less than $p_1$ is reached, and regions in which $y$ is larger than $p_2$ are avoided for $t_2$ seconds." STL is also endowed with a metric called *robustness degree* that quantifies how strongly a given trajectory satisfies an STL formula as a real number rather than just providing a *yes* or *no* answer [11], [10]. This measure enables the use of optimization methods to solve inference (e.g., [15], [18]) or formal synthesis problems (e.g., [21]) involving STL.

In this paper, we formulate two problems that enforce a desired STL specification by maximizing 1) the probability of satisfaction and 2) the expected robustness degree. One of the difficulties in solving these problems is the history-dependence of the satisfaction. For instance, if the specification requires visiting region *A* before region *B*, whether or not the system should move towards region *B* depends on whether or not it has previously visited region *A*. For LTL formulae with time-abstract semantics, this history-dependence can be broken by translating the formula to a deterministic Rabin automaton, i.e., a model that automatically takes care of the history-dependent "book-keeping", e.g., [22]. In the case of STL, such a construction is difficult due to the time-bounded semantics. We circumvent this problem by defining a fragment of STL such that the progress towards satisfaction is checked with a sufficient number of (i.e., $\tau$) state measurements. We thus define a Markov decision process (MDP), called the $\tau$-MDP, whose states correspond to the $\tau$-step history of the system and the actions are from a finite set of motion primitives.

Even though the history dependence issue can be solved by defining a $\tau$-MDP, a reinforcement learning strategy such as *Q*-learning [26] is still not applicable to maximize probability of satisfaction or expected robustness degree. In *Q*-learning, an agent tries an action, observes an immediate reward, and updates its policy to maximize the sum of rewards. However, based on the quantitative semantics of STL, the objective functions such as probability of satisfaction or expected robustness degree are not in the standard form of *Q*-learning. Thus, we propose an approximation of these objective functions such that the new synthesis problems can be solved via *Q*-learning. Moreover, we provide some performance bounds for the approximate solutions, which can be sufficiently close to the actual solutions with a proper selection of the approximation parameter. Finally, we demonstrate the performance of the proposed approach through simulation case studies.

## II. PRELIMINARIES: SIGNAL TEMPORAL LOGIC (STL)

In this paper, the desired system behavior is described by an STL fragment with the following *syntax*:

$$\begin{aligned}
\Phi &:= F_{[a,b]}\phi \mid G_{[a,b]}\phi \\
\phi &:= F_{[c,d]}\varphi \mid G_{[c,d]}\varphi \\
\varphi &:= \psi \mid \neg \varphi \mid \varphi \wedge \varphi \mid \varphi \vee \varphi,
\end{aligned} \quad (1)$$

where $a,b,c,d \in \mathbb{R}_{\geq 0}$ are finite non-negative time bounds; $\Phi$, $\phi$, and $\varphi$ are STL formulae; $\psi$ is a predicate in the form of $f(\mathbf{s}) < d$ where $\mathbf{s}: \mathbb{R}_{\geq 0} \to \mathbb{R}^n$ is a signal, $f: \mathbb{R}^n \to \mathbb{R}$ is a function, and $d \in \mathbb{R}$ is a constant. The Boolean operators $\neg$, $\wedge$, and $\vee$ are negation, conjunction (i.e., *and*), and disjunction (i.e., *or*), respectively. The temporal operators $F$ and $G$ refer to *Finally* (i.e., eventually) and *Globally* (i.e., always), respectively.

For any signal $\mathbf{s}$, let $s_t$ denote the value of $\mathbf{s}$ at time $t$ and let $(\mathbf{s},t)$ be the part of the signal that is a sequence of $s_{t'}$ for $t' \in [t,\infty)$. Accordingly, the *Boolean semantics* of STL is recursively defined as follows:

$$\begin{aligned}
(\mathbf{s},t) &\models (f(\mathbf{s}) < d) &\Leftrightarrow\;& f(s_t) < d, \\
(\mathbf{s},t) &\models \neg(f(\mathbf{s}) < d) &\Leftrightarrow\;& \neg((\mathbf{s},t) \models (f(\mathbf{s}) < d)), \\
(\mathbf{s},t) &\models \phi_1 \wedge \phi_2 &\Leftrightarrow\;& (\mathbf{s},t) \models \phi_1 \text{ and } (\mathbf{s},t) \models \phi_2, \\
(\mathbf{s},t) &\models \phi_1 \vee \phi_2 &\Leftrightarrow\;& (\mathbf{s},t) \models \phi_1 \text{ or } (\mathbf{s},t) \models \phi_2, \\
(\mathbf{s},t) &\models G_{[a,b]}\phi &\Leftrightarrow\;& (\mathbf{s},t') \models \phi \quad \forall t' \in [t+a,t+b], \\
(\mathbf{s},t) &\models F_{[a,b]}\phi &\Leftrightarrow\;& \exists t' \in [t+a,t+b] \text{ s.t. } (\mathbf{s},t') \models \phi.
\end{aligned}$$

For a signal $(\mathbf{s},0)$, i.e., the whole signal starting from time 0, satisfying $F_{[a,b]}\phi$ means that "there exists a time within $[a,b]$ such that $\phi$ will eventually be true", and satisfying $G_{[a,b]}\phi$ means that "$\phi$ is true for all times between $[a,b]$".

STL is endowed with a metric called *robustness degree* [11], [10] (also called "degree of satisfaction") that quantifies how well a given signal $\mathbf{s}$ satisfies a given formula $\Phi$. The robustness degree is calculated recursively according to the *quantitative semantics:*

$$\begin{aligned}
r(\mathbf{s},(f(\mathbf{s})<d),t) &= d - f(s_t), \\
r(\mathbf{s},\neg(f(\mathbf{s})<d),t) &= -r(\mathbf{s},(f(\mathbf{s})<d),t), \\
r(\mathbf{s},\phi_1 \wedge \phi_2,t) &= \min\left(r(\mathbf{s},\phi_1,t), r(\mathbf{s},\phi_2,t)\right), \\
r(\mathbf{s},\phi_1 \vee \phi_2,t) &= \max\left(r(\mathbf{s},\phi_1,t), r(\mathbf{s},\phi_2,t)\right), \\
r(\mathbf{s},G_{[a,b]}\phi,t) &= \min_{t' \in [t+a,t+b]} r(\mathbf{s},\phi,t'), \\
r(\mathbf{s},F_{[a,b]}\phi,t) &= \max_{t' \in [t+a,t+b]} r(\mathbf{s},\phi,t').
\end{aligned}$$

As a short-hand notation, $r(\mathbf{s},\phi)$ refers to $r(\mathbf{s},\phi,0)$ throughout the paper. Let $\varepsilon$-perturbation be a sequence of disturbances such that any signal under $\varepsilon$-perturbation stays inside the $\varepsilon$-envelope. Note that $r(\mathbf{s},\phi) = \varepsilon > 0$ means that $\mathbf{s}$ satisfies $\phi$. Moreover, the signal $\mathbf{s}$ under $\varepsilon$-perturbation still satisfies $\phi$. Similarly, $r(\mathbf{s},\phi) = \varepsilon < 0$ means that $\mathbf{s}$ violates $\phi$, and $\mathbf{s}$ under $\varepsilon$-perturbation still violates $\phi$.

As in [9], let $hrz(\phi)$ denote the *horizon length* of an STL formula $\phi$, which is the required number of samples to

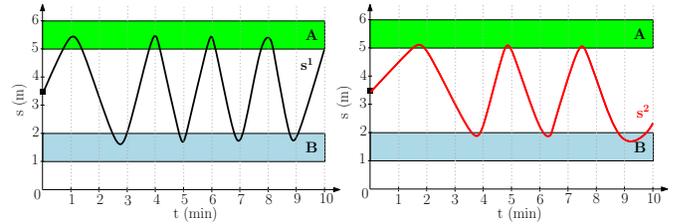

Fig. 1. The specification is "visit regions *A* and *B* every 3 minutes along a mission horizon of 10 minutes", i.e., $\Phi = G_{[0,7]}(F_{[0,3]}(s>5 \wedge s<6) \wedge F_{[0,3]}(s>1 \wedge s<2))$, which is satisfied by signal $\mathbf{s^1}$ and violated by signal $\mathbf{s^2}$.

resolve any (future or past) requirements of $\phi$. The horizon length can be computed recursively as

$$\begin{aligned}
hrz(\psi) &= 0, \\
hrz(\phi) &= b \quad \text{if } \phi = G_{[a,b]}\psi \text{ or } F_{[a,b]}\psi, \\
hrz(F_{[a,b]}\phi) &= b + hrz(\phi), \\
hrz(G_{[a,b]}\phi) &= b + hrz(\phi), \\
hrz(\neg\phi) &= hrz(\phi), \\
hrz(\phi_1 \wedge \phi_2) &= \max(hrz(\phi_1), hrz(\phi_2)), \\
hrz(\phi_1 \vee \phi_2) &= \max(hrz(\phi_1), hrz(\phi_2)),
\end{aligned}$$

where $a,b \in \mathbb{R}_{\geq 0}$, $\psi$ is a predicate, and $\phi, \phi_1, \phi_2$ are STL formulae.

**Example 1:** Consider the regions A and B illustrated in Fig. 1 and a specification as "visit regions *A* and *B* every 3 minutes along a mission horizon of 10 minutes". Note that the desired specification can be formulated in STL as

$$\begin{aligned}
\Phi &= G_{[0,7]}\phi \\
\phi &= F_{[0,3]}(s>5 \wedge s<6) \wedge F_{[0,3]}(s>1 \wedge s<2).
\end{aligned} \quad (2)$$

The horizon lengths of $\Phi$ and $\phi$ are $hrz(\Phi) = 10$ and $hrz(\phi) = 3$, respectively. Let $\psi_1 = (s>5 \wedge s<6)$ and $\psi_2 = (s>1 \wedge s<2)$. Then satisfying $\Phi$ implies satisfying $\bigwedge_{t\in[0,7]}(F_{[t,t+3]}\psi_1 \wedge F_{[t,t+3]}\psi_2)$. Let $\mathbf{s^1}$ and $\mathbf{s^2}$ be two signals as illustrated in Fig. 1. The signal $\mathbf{s^1}$ satisfies $\Phi$ because *A* and *B* are visited within $[t,t+3]$ for every $t \in [0,7]$. However, the signal $\mathbf{s^2}$ violates $\Phi$ because region *B* is not visited within $[0,3]$. Moreover, the robustness degree of $\mathbf{s}$ with respect to $\Phi$ can be computed via the quantitative semantics as follows:

$$\min_{t\in[0,7]} \min\left\{ \max_{t' \in [t,t+3]} r(\mathbf{s},\psi_1,t'), \max_{t' \in [t,t+3]} r(\mathbf{s},\psi_2,t') \right\} \quad (3)$$

Based on (3), the robustness degrees of $\mathbf{s^1}$ and $\mathbf{s^2}$ with respect to $\Phi$ are computed as $r(\mathbf{s^1},\Phi) = 0.35$ and $r(\mathbf{s^2},\Phi) = -1$ indicating that $\mathbf{s^1}$ satisfies $\Phi$ while $\mathbf{s^2}$ does not.

## III. PROBLEM FORMULATION

### A. System Model

We consider a system as a Markov decision process (MDP) $M = \langle \Sigma, A, P, R \rangle$, where $\Sigma$ denotes the state-space, $A$ is a finite set of motion primitives, $P: \Sigma \times A \times \Sigma \to [0,1]$ is a probabilistic transition relation, and $R: \Sigma \to \mathbb{R}$ is a reward function. We assume that $\Sigma$ comprises a set of partitions and each $\sigma_i \in \Sigma$ corresponds to the centroid of a partition, e.g., $\sigma_1 = (\Delta x/2, \Delta y/2)$ in Fig. 2(a). Moreover, each motion primitive $a \in A$ drives the system from the current state $\sigma_i$ to an adjacent state $\sigma_j$. Let $s_t \in \Sigma$ denote the state of a system at time $t$, and let $s_{t_1:t_2}$ denote the state trajectory of the system within $[t_1, t_2]$. Suppose that a system moves in an environment shown in Fig. 2(a), and its initial state is $s_0 = \sigma_1$. If the system visits $\sigma_3$ and returns to $\sigma_1$, its state trajectory can be written as $s_{0:2\Delta t} = \sigma_1 \sigma_3 \sigma_1$ where $\Delta t > 0$ is the discrete time step.

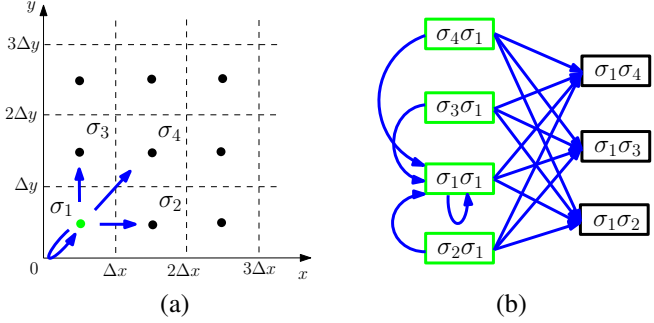

Fig. 2. (a) Discretized state-space, (b) Representation of $\sigma_1$ over $2-MDP$.

In this paper, we assume that the MDP model $M$ is already available. For more generic cases, abstractions of stochastic systems can be constructed via several methods (e.g., [16], [1], [19]). Moreover, if the system satisfies certain conditions, a discrete-time signal can be used to reason about whether or not the continuous-time signal satisfies a TL formula (e.g., [14], [13], [7]).

### B. Problem Definition

In real-world applications, many systems (e.g., robotic systems) contain uncertainty in their dynamics due to mechanical, environmental, or sensing issues. In this aspect, we consider an MDP $M$, for which the transition probability function $P$ is unknown. This means that when the system executes a motion primitive $a$ at state $s_t$, it is not certain where it will be in $t+1$, i.e., the probability distribution for $s_{t+1}$ is unknown. Then, the question becomes how to enforce an STL specification $\Phi$ to a system with unknown dynamics.

In this paper, we formulate two problems that have different objective functions to find a control policy $\pi$ enforcing the desired specification $\Phi$. In the first problem, we maximize the probability of satisfying $\Phi$, which is a commonly used objective in formal synthesis problems (e.g., [17], [19], [8]). In the second problem, we maximize the expected robustness degree with respect to $\Phi$, which has recently been used in model predictive control framework (e.g., [21], [23]).

**Problem 1 (Maximizing Probability of Satisfaction):** Let $\Phi$ be an STL specification with $hrz(\Phi) = T$. Given a stochastic model $M = \langle \Sigma, A, P, R \rangle$ with unknown $P$, known reward function $R$, and an initial partial state trajectory $s_{0:\tau}$ for some $\tau \in [0, T)$, find a control policy

$$\pi_1^* = \arg\max_\pi Pr^\pi[s_{0:T} \models \Phi] \quad (4)$$

where $Pr^\pi[s_{0:T} \models \Phi]$ is the probability of $s_{0:T}$ satisfying $\Phi$ under policy $\pi$.

**Problem 2 (Maximizing Expected Robustness Degree):** Let $\Phi$ be an STL specification with $hrz(\Phi) = T$. Given a stochastic model $M = \langle \Sigma, A, P, R \rangle$ with unknown $P$, known reward function $R$, and an initial partial state trajectory $s_{0:\tau}$ for some $\tau \in [0, T)$, find a control policy

$$\pi_2^* = \arg\max_\pi E^\pi[r(s_{0:T}, \Phi)] \quad (5)$$

where $E^\pi[r(s_{0:T}, \Phi)]$ is the expected robustness degree of $s_{0:T}$ with respect to $\Phi$ under policy $\pi$.

## IV. CONTROL SYNTHESIS VIA $Q$-LEARNING

For systems with unknown stochastic dynamics, reinforcement learning can be used to design optimal control policies, that is, the system learns how to take actions by trial and error interactions with the environment [24]. In this paper, we use the $Q$-learning algorithm that is briefly presented in the first sub-section. Then, we discuss that Problems 1 and 2 are not in the standard form to apply this algorithm. Finally, we present the main contribution of this paper, i.e., the approximation of STL synthesis problems that can be solved via $Q$-learning.

### A. Q-learning

$Q$-learning is a model-free reinforcement learning method [26], which can be used to find the optimal policy for a finite MDP. In particular, the objective of an agent at state $s_t$ is to maximize $V(s_t)$, its expected (discounted) reward in finite or infinite horizon, i.e.,

$$E\left[\sum_{k=0}^{T} r(s_{k+t+1})\right] \quad \text{or} \quad E\left[\sum_{k=0}^{\infty} \gamma^k r(s_{k+t+1})\right], \quad (6)$$

where $r(s)$ is the reward obtained at state $s$, and $\gamma$ is the discount factor. Also, $V^*(s) = \max_a Q^*(s,a)$, where $Q^*(s,a)$ is the optimal Q-function for every state-action pair $(s,a)$.

Starting from state $s$, the system chooses an action $a$, which takes it to state $s'$ and results in a reward $r$. Then, the $Q$-learning rule is defined as follows:

$$Q(s,a) := (1-\alpha)Q(s,a) + \alpha[r + \gamma \max_{a^* \in A} Q(s', a^*)], \quad (7)$$

where $\gamma \in (0,1)$ is the discount factor and $\alpha \in (0,1]$ is the learning rate. Accordingly, if each action $a \in A$ is repetitively implemented in each state $s \in \Sigma$ for infinite number of times and $\alpha$ decays appropriately, then $Q$ converges to $Q^*$ with probability 1 (see Theorem 4.1). Thus, we can find the optimal policy $\pi^* : \Sigma \to A$ as $\pi^* = \arg\max_a Q^*(s,a)$. Algorithm 1 shows the steps of $Q$-learning.

**Algorithm 1:** *Q*-learning [24]

*Input:* $s$ - current state
*Output:* $\pi$ - control policy maximizing the sum of (discounted) rewards

1: **initialization:** Arbitrary $Q(s,a)$ and $\pi$;
2: **for** $k = 1 : N_{episode}$
3:    Initialize $s$
3:    **for** $t = 1 : T$
4:        Select an action $a$ (via $\varepsilon$-greedy or $\pi$);
5:        Take action $a$, observe $r$ and $s'$;
6:        $Q(s,a) \leftarrow (1-\alpha_k)Q(s,a) + \alpha_k[r + \gamma\max_{a'} Q(s',a')]$;
7:        $\pi(s) \leftarrow \arg\max_a Q(s,a)$;
8:        $s \leftarrow s'$;
9:    **end for**
10: **end for**

**Theorem 4.1:** [25] Given a finite MDP, $M = \langle S,A,P,R \rangle$, let $Q^*(s,a)$ be the optimal Q-function for every pair of $(s,a)$. Consider the *Q*-learning algorithm with the update rule

$$Q_{k+1}(s,a) = (1-\alpha_k)Q_k(s,a) + \alpha_k[r + \gamma\max_{a^* \in A} Q_k(s',a^*)],$$

where the discount factor $\gamma \in (0,1)$, $\alpha_k$ satisfies $\sum_k \alpha_k = \infty$ and $\sum_k \alpha_k^2 < \infty$. Then $Q_k(s,a)$ converges to $Q^*(s,a)$ with probability 1 as $k \to \infty$.

### B. Q-learning and Formal Synthesis

There are several reasons why one cannot directly use *Q*-learning in Problems 1 and 2. First of all, the action selection at each time step cannot depend on only the current state as in *Q*-learning. For example, consider a specification $\Phi_1 = F_{[0,T]}\psi$ where $\psi = x > 3$. Satisfying $\Phi_1$ implies that visiting the desired region at least one time in $[0,T]$. Let the current state be $s_t := x = 2$ and assume that the desired region is not visited before $t$. Note that if $t = T - 1$, then the action selection via the optimal policy leads the agent to maximally approach the desired region. However, if $t = 0$, then the optimal policy may result in an action that drives the agent further away from the desired region (while ensuring to eventually satisfy $\psi$). Thus, the optimal policies may not necessarily be the same if the same state is occupied but the remaining mission horizons are different. Moreover, if $\Phi$ involves a nested temporal operator, the policy should also take into account a sufficient length of state history in addition to the current state and the remaining mission horizon. For example, consider $\Phi_2 = F_{[0,T]}G_{[0,\tau]}\psi$. Note that $\Phi_2$ implies that the agent should eventually enter the desired region in $[0,T]$ and stay there for $\tau$ time steps. Similarly, let the current state be $s_t = 2$. The action selection at $t$ depends on the state history $s_{t-\tau:t}$ and the remaining mission horizon. Thus, the policies in Problems 1 and 2 should be defined as $\pi : \Sigma^\tau \times \mathbb{N}_{\geq 0} \to A$ where $\Sigma^\tau = \Sigma \times \cdots \times \Sigma$ for $\tau$ times.

Secondly, one can not directly apply *Q*-learning because an agent trying to optimize (4) or (5) does not have an immediate reward after taking an action. Consider a specification $\Phi$ such that $hrz(\Phi) = T$. Accordingly, both satisfaction and the robustness degree can be computed over a $T$-length trajectory (i.e., these measures are undefined for partial trajectories having a length smaller than $T$). For example, consider an agent trying to satisfy $\Phi_1 = F_{[0,T]}\psi$. Then, the objective function in Problem 2 can be written as

$$\max_\pi E^\pi \Big[\max\big(r(s_{0:T},\psi,0),\ldots,r(s_{0:T},\psi,T)\big)\Big]. \quad (8)$$

Hence, the objective functions in (4) or (5) are not in the standard form of *Q*-learning as in (6).

### C. Proposed Approach

In this paper, we approximate the synthesis problems in (4) and (5) such that one can use the *Q*-learning algorithm to find the optimal policy. The overview of the proposed method is: 1) for any STL formula (i.e., $G_{[0,T]}\phi$ or $F_{[0,T]}\phi$), redefine the state-space as $\Sigma^\tau$ where $\tau$ is a function of $hrz(\phi)$; 2) redefine the objective function such that an agent observes an immediate reward after taking each action and the remaining mission horizon can be eliminated in the policy design. After executing these steps, we will show that one can use the *Q*-learning algorithm to find the optimal policy $\pi^* : \Sigma^\tau \to A$.

Let $\Phi$ be $G_{[0,T]}\phi$ or $F_{[0,T]}\phi$, where $\Phi$ and $\phi$ are STL formulae with the syntax in (1). Let the horizon length of $\phi$ be $hrz(\phi) = \tau$. Then, we denote the $\tau$-state of the agent at time $t$ by $s_t^\tau$, which is the $\tau$-horizon trajectory involving the current state and the most recent $\tau - 1$ past states, i.e., $s_t^\tau = s_{t-\tau+1:t}$. By considering all $\tau$-states of the agent, we remodel the agent as a $\tau$-MDP.

**Definition 1 ($\tau$-MDP):** Given an MDP $M = (\Sigma,A,P,R)$ and $\tau \in \mathbb{N}_{>0}$, a $\tau$-MDP is a tuple $M^\tau = (\Sigma^\tau, A, P^\tau, R^\tau)$, where

- $\Sigma^\tau \subseteq (\Sigma \cup \varepsilon)^\tau$ is the set of finite states, where $\varepsilon$ is the empty string. Each state $\sigma^\tau \in \Sigma^\tau$ corresponds to a $\tau$-horizon (or shorter) path on $\Sigma$. Shorter paths of length $n < \tau$ (representing the case in which the system has not yet evolved for $\tau$ time steps) have $\varepsilon$ prepended $\tau - n$ times.

- $P^\tau : \Sigma^\tau \times A \times \Sigma^\tau \to [0,1]$ is a probabilistic transition relation. Let $\sigma_i^\tau = \sigma_a\sigma_b\ldots\sigma_c\sigma_d$ and $\sigma_j^\tau = \sigma_e\ldots\sigma_f\sigma_g$. $P^\tau(\sigma_i^\tau,a,\sigma_j^\tau) > 0$ if and only if $P(\sigma_d,a,\sigma_g) \in [0,1]$ and for $\tau > 1$ the first $\tau - 1$ elements of $\sigma_j^\tau$ are equal to the last $\tau - 1$ elements of $\sigma_i^\tau$ (i.e., $\sigma_e\ldots\sigma_f = \sigma_b\ldots\sigma_d$).

- $R^\tau : \Sigma^\tau \to \mathbb{R}$ is a reward function.

For instance, the highlighted state $\sigma_1$ in Fig. 2(a) corresponds to four $\tau$-states for $\tau = 2$ as illustrated in Fig. 2(b).

For any $\Phi = F_{[0,T]}\phi$ or $\Phi = G_{[0,T]}\phi$, $\tau$ can be computed as follows:

$$\tau = \left\lceil \frac{hrz(\phi)}{\Delta t} \right\rceil + 1 \quad (9)$$

where $\Delta t$ is the time step and $\lceil . \rceil$ is the ceiling function, i.e., $\lceil x \rceil$ is the smallest integer not less than $x \in \mathbb{R}$.

**Remark 1:** If $\Phi$ does not have nested temporal operators, then $hrz(\phi) = 0$ and $\tau = 1$ as a consequence. As such, $M^\tau = M$ for $\tau = 1$.

For any state trajectory $s_{0:T}$, we can write the corresponding $\tau$-state trajectory as $s_{\tau-1:T}^\tau = s_{\tau-1}^\tau \ldots s_T^\tau$ where each $s_t^\tau := s_{t-\tau+1:t}$ for $\tau - 1 \leq t \leq T$. Moreover, for each $\tau$-state $s_t^\tau$, we can compute the corresponding robustness degree with respect to $\phi$. Accordingly, the robustness degree of $s_{0:T}$ with respect to $\Phi$ can be written in terms of $\tau$-states as

$$r(s_{0:T}, \Phi) = \begin{cases} \max\left(r(s^\tau_{\tau-1}, \phi), \ldots, r(s^\tau_T, \phi)\right), & \text{if } \Phi = F_{[0,T]}\phi \\ \min\left(r(s^\tau_{\tau-1}, \phi), \ldots, r(s^\tau_T, \phi)\right), & \text{if } \Phi = G_{[0,T]}\phi \end{cases} \quad (10)$$

Note that plugging (10) into (5) makes the objective in Problem 2 as follows:

$$\max_\pi E^\pi[r(s_{0:T}, \Phi)] = \begin{cases} \max_\pi E^\pi\left[\max_{\tau-1 \leq t \leq T} (r(s^\tau_t, \phi))\right], & \text{if } \Phi = F_{[0,T]}\phi \\ \max_\pi E^\pi\left[\min_{\tau-1 \leq t \leq T} (r(s^\tau_t, \phi))\right], & \text{if } \Phi = G_{[0,T]}\phi \end{cases} \quad (11)$$

Since $Q$-learning cannot be used for cases like (11), we propose to use the *log-sum-exp* [6] approximation of the maximum function to represent the objective as a sum of rewards, i.e.,

$$\max(x_1, \ldots, x_n) \sim \frac{1}{\beta} \log \sum_{i=1}^n e^{\beta x_i}, \quad (12)$$

where $\beta > 0$ is a constant and

$$\max(x_1, \ldots, x_n) \leq \frac{1}{\beta} \log \sum_{i=1}^n e^{\beta x_i} \leq \max(x_1, \ldots, x_n) + \frac{1}{\beta} \log n, \quad (13)$$

meaning that $\frac{1}{\beta} \log \sum_{i=1}^n e^{\beta x_i}$ can approximate $\max(x_1, \ldots, x_n)$ with arbitrary accuracy by selecting a large $\beta$. Based on (12), the equation in (11) can be approximated as

$$\max_\pi E^\pi[r(s_{0:T}, \Phi)] \sim \begin{cases} \max_\pi E^\pi\left[\frac{1}{\beta} \log \sum_{t=\tau-1}^T e^{\beta r(s^\tau_t, \phi)}\right], & \text{if } \Phi = F_{[0,T]}\phi \\ \max_\pi E^\pi\left[-\frac{1}{\beta} \log \sum_{t=\tau-1}^T e^{-\beta r(s^\tau_t, \phi)}\right], & \text{if } \Phi = G_{[0,T]}\phi \end{cases} \quad (14)$$

Similarly, maximizing the probability of satisfying $\Phi$ can be written as

$$\max_\pi Pr^\pi[s_{0:T} \models \Phi] = \max_\pi E^\pi\left[I(r(s_{0:T}, \Phi))\right] \quad (15)$$

where $I(.)$ is the indicator function defined as

$$I(x) = \begin{cases} 1, & \text{if } x \geq 0 \\ 0, & \text{otherwise.} \end{cases} \quad (16)$$

Since $I(\max(x_1, \ldots, x_n)) = \max(I(x_1), \ldots, I(x_n))$ (or $I(\min(x_1, \ldots, x_n)) = \min(I(x_1), \ldots, I(x_n))$), plugging (10) into (15) makes the objective in Problem 1 as follows:

$$\max_\pi Pr^\pi[s_{0:T} \models \Phi] = \begin{cases} \max_\pi E^\pi\left[\max_{\tau-1 \leq t \leq T} I(r(s^\tau_t, \phi))\right], & \text{if } \Phi = F_{[0,T]}\phi \\ \max_\pi E^\pi\left[\min_{\tau-1 \leq t \leq T} I(r(s^\tau_t, \phi))\right], & \text{if } \Phi = G_{[0,T]}\phi \end{cases} \quad (17)$$

Based on (12), the equation in (17) can be approximated as

$$\max_\pi Pr^\pi[s_{0:T} \models \Phi] \sim \begin{cases} \max_\pi E^\pi\left[\frac{1}{\beta} \log \sum_{t=\tau-1}^T e^{\beta I(r(s^\tau_t, \phi))}\right], & \text{if } \Phi = F_{[0,T]}\phi \\ \max_\pi E^\pi\left[-\frac{1}{\beta} \log \sum_{t=\tau-1}^T e^{-\beta I(r(s^\tau_t, \phi))}\right], & \text{if } \Phi = G_{[0,T]}\phi \end{cases} \quad (18)$$

**Problem 1A (Max. Approx. Probability of Satisfaction):** Let $\Phi$ and $\phi$ be STL formulae with the syntax in (1) such that $\Phi = F_{[0,.]}\phi$ or $\Phi = G_{[0,.]}\phi$. Let $hrz(\Phi) = T$. Given an unknown MDP $M$, let $M^\tau = \langle \Sigma^\tau, A, P^\tau, R^\tau \rangle$ be the $\tau$-MDP where $\tau$ is computed as in (9). Assume that the initial $\tau$-state $s^\tau_{\tau-1} = s_{0:\tau-1}$ is given and $\beta > 0$. Find a control policy $\pi^*_{1A} : \Sigma^\tau \to A$ such that

$$\pi^*_{1A} = \begin{cases} \arg\max_\pi E^\pi\left[\sum_{t=\tau-1}^T e^{\beta I(r(s^\tau_t, \phi))}\right], & \text{if } \Phi = F_{[0,T]}\phi \\ \arg\max_\pi E^\pi\left[-\sum_{t=\tau-1}^T e^{-\beta I(r(s^\tau_t, \phi))}\right], & \text{if } \Phi = G_{[0,T]}\phi \end{cases} \quad (19)$$

**Problem 2A (Max. Expected Approx. Robustness Degree):** Let $\Phi$ and $\phi$ be STL formulae with the syntax in (1) such that $\Phi = F_{[0,.]}\phi$ or $\Phi = G_{[0,.]}\phi$. Let $hrz(\Phi) = T$. Given an unknown MDP $M$, let $M^\tau = \langle \Sigma^\tau, A, P^\tau, R^\tau \rangle$ be the $\tau$-MDP where $\tau$ is computed as in (9). Assume that the initial $\tau$-state $s^\tau_{\tau-1} = s_{0:\tau-1}$ is given and $\beta > 0$. Find a control policy $\pi^*_{2A} : \Sigma^\tau \to A$ such that

$$\pi^*_{2A} = \begin{cases} \arg\max_\pi E^\pi\left[\sum_{t=\tau-1}^T e^{\beta r(s^\tau_t, \phi)}\right], & \text{if } \Phi = F_{[0,T]}\phi \\ \arg\max_\pi E^\pi\left[-\sum_{t=\tau-1}^T e^{-\beta r(s^\tau_t, \phi)}\right], & \text{if } \Phi = G_{[0,T]}\phi \end{cases} \quad (20)$$

**Theorem 4.2:** Let $\Phi$ and $\phi$ be STL formulae with the syntax in (1) such that $\Phi = F_{[0,.]}\phi$ or $\Phi = G_{[0,.]}\phi$. Let $hrz(\Phi) = T$. Assume that a partial state trajectory $s_{0:\tau-1}$ is initially given where $\tau$ is computed as in (9). For some $\beta > 0$ and $\Delta t = 1^1$, let $\pi^*_1, \pi^*_2, \pi^*_{1A}, \pi^*_{2A}$ be the optimal policies obtained by solving Problems 1, 2, 1A, 2A, respectively. Then,

$$Pr^{\pi^*_1}[s_{0:T} \models \Phi] - \frac{1}{\beta} \log(T - \tau + 2) \leq Pr^{\pi^*_{1A}}[s_{0:T} \models \Phi]$$

$$E^{\pi^*_2}[r(s_{0:T}, \Phi)] - \frac{1}{\beta} \log(T - \tau + 2) \leq E^{\pi^*_{2A}}[r(s_{0:T}, \Phi)]$$

*Proof:* First, we will show that solving (19) is equivalent to solving the right hand-side of (18). Let $\mathbf{s}^\tau = s^\tau_{\tau-1:T}$ and

$$g(\mathbf{s}^\tau) = \begin{cases} \sum_{t=\tau-1}^T e^{\beta I(r(s^\tau_t, \phi))}, & \text{if } \Phi = F_{[0,T]}\phi \\ -\sum_{t=\tau-1}^T e^{-\beta I(r(s^\tau_t, \phi))}, & \text{if } \Phi = G_{[0,T]}\phi \end{cases} \quad (21)$$

Since $\log(.)$ is a strictly monotonic function and $1/\beta$ is a constant,

$$\arg\max_\pi E^\pi\left[g(\mathbf{s}^\tau)\right] \Leftrightarrow \arg\max_\pi E^\pi\left[\frac{1}{\beta} \log g(\mathbf{s}^\tau)\right]. \quad (22)$$

In other words, $\pi^*_{1A}$ is also the optimal policy for the right hand side of (18). Following the similar steps, we can show that solving (20) is equivalent to solving the right hand-side

---

[1]$\Delta t = 1$ is selected due to clarity in presentation, but it can be any time step.

of (14), thus $\pi_{2A}^*$ is also the optimal policy for the right hand side of (14).

Note that any $\tau$-state trajectory $\mathbf{s}^\tau$ implies a state trajectory $\mathbf{s} = s_{0:T}$. Let $\Pi$ be the set of policies. Starting from $s_{0:\tau-1}$ (i.e., initially given partial state trajectory), any $\pi \in \Pi$ induces a set of trajectories. Then, based on (13),

$$E^\pi[g(\mathbf{s}^\tau)] \leq Pr^\pi[\mathbf{s} \models \Phi] + \frac{1}{\beta}\log(T-\tau+2) \quad (23)$$

where $T-\tau+2$ is the total length of the $\tau$-state trajectory (i.e., $\mathbf{s}^\tau = s_{\tau-1}^\tau s_\tau^\tau \ldots s_T^\tau$). The equation in (23) implies that the approximation function can over-evaluate the set of trajectories obtained by a policy $\pi$ at most $\frac{1}{\beta}\log(T-\tau+2)$. Hence, $\pi_{1A}^*$ can result in a sub-optimal performance that is at most $\frac{1}{\beta}\log(T-\tau+2)$ away from the performance obtained by $\pi_1^*$. Following the same steps, we can show that $\pi_{2A}^*$ results in a sub-optimal performance that is at most $\frac{1}{\beta}\log(T-\tau+2)$ away from the performance obtained by $\pi_2^*$. ∎

In the following proposition, we show that $Q$-learning can be used to solve Problems 1A and 2A.

**Proposition 4.3:** Let $\Phi$ and $\phi$ be STL formulae such that $\Phi = F_{[0,.]}\phi$ or $\Phi = G_{[0,.]}\phi$. Given a finite MDP $M$, let $M^\tau = \langle \Sigma^\tau, A, P^\tau, R^\tau \rangle$ be the $\tau$-MDP where $\tau$ is computed as in (9) and $Q^*(s^\tau, a)$ is the optimal Q-function for every pair of $(s^\tau, a)$. Consider the $Q$-learning algorithm with the following update rule

$$Q_{k+1}(s_i^\tau, a) = (1-\alpha_k)Q_k(s_i^\tau, a) + \alpha_k[R + \gamma \max_{a^* \in A} Q_k(s_j^\tau, a^*)],$$

where $s_j^\tau$ is the resulting state by taking action $a$ at $s_i^\tau$, $\gamma \in (0,1)$, $\alpha_k$ satisfies $\sum_k \alpha_k = \infty$ and $\sum_k \alpha_k^2 < \infty$, and for some $\beta > 0$, the immediate reward $R$ obtained at $s_j^\tau$ is defined as

$$R = \begin{cases} e^{\beta I(r(s_j^\tau, \phi))}, & \text{if Problem 1A with } \Phi = F_{[0,T]}\phi \\ -e^{-\beta I(r(s_j^\tau, \phi))}, & \text{if Problem 1A with } \Phi = G_{[0,T]}\phi \\ e^{\beta r(s_j^\tau, \phi)}, & \text{if Problem 2A with } \Phi = F_{[0,T]}\phi \\ -e^{-\beta r(s_j^\tau, \phi)}, & \text{if Problem 2A with } \Phi = G_{[0,T]}\phi \end{cases} \quad (24)$$

Then $Q_k(s^\tau, a)$ converges to $Q^*(s^\tau, a)$ with probability 1 as $k \to \infty$.

*Proof:* The proof follows from Theorem 4.1. ∎

**Remark 2:** When $Q$-learning is used to maximize the discounted versions of (19) and (20) as in Proposition 4.3, the resulting performance gap can be derived from (13) as

$$\max(x_1,\ldots,x_n) + \frac{1}{\beta}\log \gamma^n \leq \frac{1}{\beta}\log \sum_{i=1}^n \gamma^i e^{\beta x_i} \leq \max(x_1,\ldots,x_n) + \frac{1}{\beta}\log n.$$

Hence, the result of Theorem 4.2 via $Q$-learning can be extended as

$$Pr^{\pi_1^*}[s_{0:T} \models \Phi] - \frac{1}{\beta}\max\left(\log(T-\tau+2) - \log\gamma^n\right) \leq Pr^{\pi_{1A}^*}[s_{0:T} \models \Phi]$$

$$E^{\pi_2^*}[r(s_{0:T}, \Phi)] - \frac{1}{\beta}\max\left(\log(T-\tau+2) - \log\gamma^n\right) \leq E^{\pi_{2A}^*}[r(s_{0:T}, \Phi)].$$

Consequently, selecting $\gamma$ close to 1 and arbitrarily large selection of $\beta$ significantly reduces the performance gap between the solutions obtained via Problems 1 and 1A (or 2 and 2A). However, larger values of $\beta$ would increase the maximum reward hence would reduce the convergence rate in $Q$-learning [20].

## V. SIMULATION RESULTS

In the following case studies, we consider a single agent moving in a discretized environment. The set of motion primitives at each state is $A = \{N, NW, W, SW, S, SE, E, NE, stay\}$. We model the motion uncertainty as in Fig. 3 where, for any selected feasible action in $A$, the agent follows the corresponding blue arrow with probability 0.93 or a red arrow with probability 0.023. Moreover, the resulting state after taking an infeasible action (i.e., the agent is next to a boundary and tries to move towards it) is the current state. All simulations were implemented in MATLAB and performed on a PC with a 2.8 GHz processor and 8.0 GB RAM.

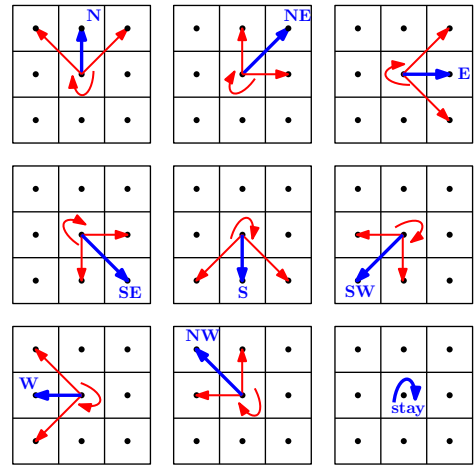

Fig. 3. The motion uncertainty (as red arrows) for a particular action (blue arrow).

### A. Case Study 1: Reachability

In this case study, the initial state of the agent is $s_0 = (1.5, 1.5)$ as shown in Fig. 4. We consider an STL formula defined over the environment as

$$\Phi_1 = F_{[0,7]}(x > 4 \land y > 4), \quad (25)$$

which expresses "eventually visit the desired region within $[0,7]$. Note that $\Phi_1 = F_{[0,7]}\phi$ where $\phi = (x > 4 \land y > 4)$ and $hrz(\phi) = 0$. Moreover, we choose $\Delta t = 1$, thus $\tau = 1$ from (9). The state-spaces of the system based on Fig. 4 are $|\Sigma| = 36$ and $|\Sigma^\tau| = |\Sigma|$ since $\tau = 1$.

To implement the $Q$-learning algorithm, the number of episodes is chosen as 1700 (i.e., $1 \leq k \leq 1700$), and we use the parameters $\beta = 50$, $\gamma = 0.9999$, and $\alpha_k = 0.95^k$. After 1700 trainings (episodes), which took approximately 1 minute for each problem, the resulting policies $\pi_{1A}^*$ and $\pi_{2A}^*$ are used to generate 1000 trajectories, which lead to

$$E^{\pi_{1A}^*}[r(s_{0:7}, \Phi_1)] = 0.523, \quad Pr^{\pi_{1A}^*}[s_{0:7} \models \Phi_1] = 0.999,$$
$$E^{\pi_{2A}^*}[r(s_{0:7}, \Phi_1)] = 1.497, \quad Pr^{\pi_{2A}^*}[s_{0:7} \models \Phi_1] = 1.000.$$

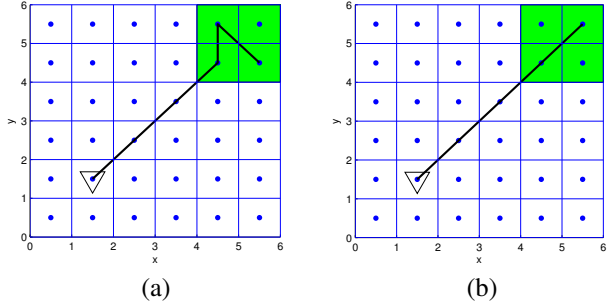

Fig. 4. The environment and the desired region in case study 1 for which a sample trajectory by (a) $\pi_{1A}^*$ and (b) $\pi_{2A}^*$.

Sample trajectories generated by $\pi_{1A}^*$ and $\pi_{2A}^*$ are displayed in Fig. 4 (a) and (b), respectively. $\Phi_1$ is satisfied with very high probability in the first case and with probability 1 in the second case. The trajectories via maximizing the expected robustness degree tend to reach the deepest state (i.e., having the maximum robustness degree with respect to $\phi$) in the desired region.

*B. Case Study 2: Repeated Satisfiability*

In the second case study, we consider an agent moving in an environment illustrated in Fig. 5(a). We consider an STL formula defined over the environment as

$$\Phi_2 = G_{[0,12]}\big(F_{[0,2]}(region\ A) \wedge F_{[0,2]}(region\ B)\big) \qquad (26)$$

where *region A* represents $x > 1 \wedge x < 2 \wedge y > 3 \wedge y < 4$ and *region B* represents $x > 2 \wedge x < 3 \wedge y > 2 \wedge y < 3$. Note that $\Phi_2$ expresses the following: "for all $t \in [0,12]$, eventually visit *region A* every $[t, t+2]$ and eventually visit *region B* every $[t, t+2]$". Note that $\Phi_2 = G_{[0,12]}\phi$ where $\phi = F_{[0,2]}(region\ A) \wedge F_{[0,2]}(region\ B)$ and $hrz(\phi) = 2$. Assuming that $\Delta t = 1$, $\tau = 3$ based on (9).

In this case study, the sizes of the state-spaces are $|\Sigma| = 16$ and $|\Sigma^\tau| = 676^2$ for $\tau = 3$. To implement the *Q*-learning algorithm, the number of episodes is chosen as 2000 (i.e., $1 \le k \le 2000$), and we use the parameters $\beta = 50$, $\gamma = 0.9999$, and $\alpha_k = 0.95^k$. After 2000 trainings, which took approximately 6 minutes for each problem, the resulting policies $\pi_{1A}^*$ and $\pi_{2A}^*$ are used to generate 500 trajectories, which lead to

$E^{\pi_{1A}^*}[r(s_{0:14}, \Phi_2)] = 0.084, \quad Pr^{\pi_{1A}^*}[s_{0:14} \models \Phi_2] = 0.732,$
$E^{\pi_{2A}^*}[r(s_{0:14}, \Phi_2)] = 0.422, \quad Pr^{\pi_{2A}^*}[s_{0:14} \models \Phi_2] = 0.936.$

Sample trajectories generated by $\pi_{1A}^*$ and $\pi_{2A}^*$ are displayed in Fig. 5 (b) and (c), respectively.

In this case study, the performances obtained by maximizing probability of satisfaction and expected robustness degree are different from each other. The discrepancy between the two solutions can be explained by what happens when trajectories almost satisfy $\Phi_2$. While solving Problem 1A, if a $\tau$-state slightly violating or strongly violating $\phi$ (i.e., a partial trajectory almost oscillating or not oscillating between the regions *A* and *B* in two seconds) is encountered, the overall reward in both cases will be the same. On the other hand, while solving Problem 2A, the policy producing the slightly violating $\tau$-state (i.e., almost oscillatory partial trajectory) will be reinforced much more strongly than an arbitrary policy as the resulting robustness degree is larger. Since the robustness degree gives "partial credit" for trajectories that are close to satisfaction, the *Q*-learning algorithm performs a directed search to find policies that satisfy the formula. Since probability maximization gives no partial credit, the *Q*-learning algorithm is essentially performing a random search until it encounters a trajectory that satisfies the given formula. Therefore, if the family of policies that satisfy the formula with positive probability is small, it will on average take the *Q*-learning algorithm solving Problem 1A a longer time to converge to a solution that enforces formula satisfaction.

[2]This indicates that there are 676 partial trajectories with 3 states in the scenario illustrated in in Fig. 5(a), and it is computed by taking into account the admissible 2 transitions at each state in Fig. 5(a).

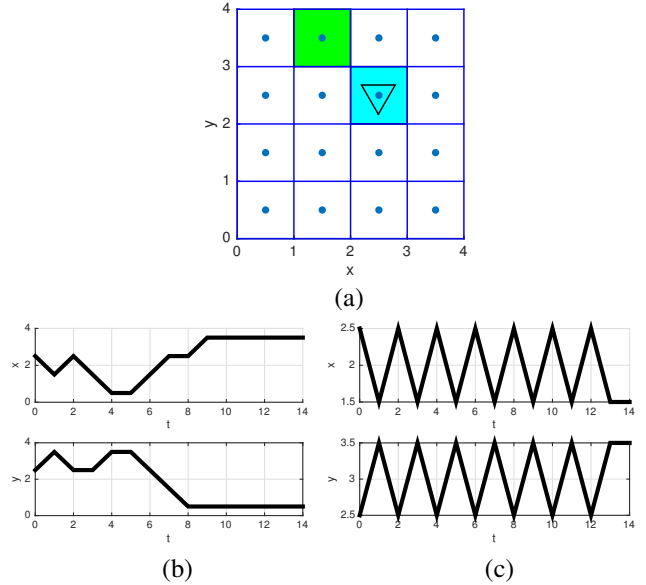

Fig. 5. (a) The initial state and the desired regions in case study 2 for which a sample trajectory by (b) $\pi_{1A}^*$ and (c) $\pi_{2A}^*$.

## VI. CONCLUSIONS AND FUTURE WORK

We considered a system which is modeled as an MDP with unknown transition probabilities and is required to satisfy a complex task given as an STL formula. To find a control policy enforcing the desired STL formula, we addressed two problems maximizing 1) the probability of satisfaction, and 2) the expected robustness degree, i.e., a measure quantifying the quality of satisfaction. One way to learn optimal policies for unknown stochastic MDPs is via *Q*-learning, where an agent receives a reward after each action; the objective is maximizing the sum of rewards; and the action selection depends only on the current state. However, the problems maximizing the probability of satisfaction and the expected robustness degree do not have the aforementioned properties.

In this paper, we proposed an approximation of STL synthesis problems that can be solved via *Q*-learning. The proposed method is based on 1) remodeling the system as a

$\tau$-MDP where each state corresponds to a $\tau$-length trajectory and $\tau$ is computed based on the given STL formula, 2) approximating the probability of satisfaction and expected robustness degree such that the new objective functions are in the form of sum of rewards. We also showed that the polices computed by the proposed method can be sufficiently close to the policies of the original problems when the approximation parameter is selected properly. Finally, we demonstrated the performance of the proposed method on some case studies, and we observed that after the same number of training, the resulting policy by maximizing the expected robustness degree performs better than the resulting policy by maximizing the probability of satisfaction. Future research includes incorporating complexity reduction techniques for faster convergence to optimal policies and extending this work for multi-agent systems.